# Chiral charge density wave and backscattering-immune orbital texture in monolayer 1$T$-TiTe$_2$

*Mingqiang Ren[1,2,†], Fangjun Cheng[1,†], Yufei Zhao[2,3,†], Mingqiang Gu[2], Qiangjun Cheng[1], Binghai Yan[3], Qihang Liu[2,\*], Xucun Ma[1,4,\*], Qikun Xue[1,2,4,5,\*], Can-Li Song[1,4,\*]*

[1]State Key Laboratory of Low-Dimensional Quantum Physics, Department of Physics, Tsinghua University, Beijing 100084, China.

[2]Shenzhen Institute for Quantum Science and Engineering and Department of Physics, Southern University of Science and Technology, Shenzhen 518055, China.

[3]Department of Condensed Matter Physics, Weizmann Institute of Science, Rehovot 7610001, Israel.

[4]Frontier Science Center for Quantum Information, Beijing 100084, China.

[5]Beijing Academy of Quantum Information Sciences, Beijing 100193, China.

[†]These authors contributed equally to this work.

Corresponding*: clsong07@mail.tsinghua.edu.cn, liuqh@sustech.edu.cn, xucunma@mail.tsinghua.edu.cn, qkxue@mail.tsinghua.edu.cn

**ABSTRACT:** Non-trivial electronic states are attracting intense attention in low-dimensional physics. Though chirality has been identified in charge states with a scalar order parameter, its intertwining with charge density waves (CDW), film thickness and the impact on the electronic behaviors remain less well understood. Here, using scanning tunneling microscopy, we report a 2 × 2 chiral CDW as well as a strong suppression of the Te-$5p$ hole-band backscattering in monolayer 1$T$-TiTe$_2$. These exotic characters vanish in bilayer TiTe$_2$ with a non-CDW state. Theoretical calculations approve that chirality comes from a helical stacking of the triple-$q$ CDW components and therefore can persist at the two-dimensional limit. Furthermore, the chirality renders the Te-$5p$ bands an unconventional orbital texture that prohibits electron backscattering. Our study establishes TiTe$_2$ as a promising playground for manipulating the chiral ground states at the monolayer limit and provides a novel path to engineer electronic properties from an orbital degree.

Manipulating and engineering the non-trivial order of charge or spin states at low-dimension, such as chirality, helicity, etc., provide newfangled paths to realize novel quantum phenomena and next-generation applications. Thanks to the vector nature, spin-ordered or spin-momentum locked states are prone to exhibit chiral or helical patterns.[1-4] However, a charge ordered state characterized by a scalar order parameter rarely manifests chiral or helical features. Recently, such state, the chiral CDW was experimentally identified in certain layered transition-metal dichalcogenides (TMDCs),[5-8] unconventional cuprate superconductors,[9] Weyl semimetals,[10] and kagome materials.[11,12] The chiral CDW phase as a spontaneous mirror symmetry breaking usually underlies the intertwining among orbital order, electron correlation and non-trivial band topology. Extensive efforts have been dedicated to directly image and manipulate the chiral CDW state,[5-17] enabling the emergence of chiral superconductivity,[13,14] ferroelectricity[15] and nonlinear optical activity.[16,17]

To date, the microscopic mechanism of the chiral CDW has not been fully understood. Unlike a conventional CDW described by one or multiple charge modulation wave vectors $q$ with the same amplitude and phase, the chiral CDW is often proposed to manifest a unique helical stacking of inequivalent $q$ vectors in a CDW unit cell (known as "triple-$q$" theory), generating clockwise or anticlockwise rotation of their intensity and thus breaking spatial inversion symmetry with an axial vector,[18,19] In both 1$T$-TiSe$_2$ and kagome materials,[19,20] a sequential phase difference between the triple-$q$ CDW vectors is expected to drive the chiral charge modulations along the layer stacking direction, and then supports a chiral CDW phase in the bulk or multilayers. To clarify the interplay between chirality and the collective CDW mode, it is therefore crucial to engineer and investigate the chiral CDW state at the two-dimensional (2D) limit.

Recently, a commensurate 2 × 2 CDW has been observed in monolayer TiTe$_2$, while no CDW instability emerges in its bulk and multilayer form.[21-28] As an isostructural sister compound to the first discovered chiral-CDW material 1$T$-TiSe$_2$, TiTe$_2$ serves as a potential platform to study chiral CDW at the 2D limit.[19] Furthermore, if chirality exists, the striking contrast between TiTe$_2$ monolayer and multilayers enables a comparative study on the chirality-induced physical effect, such as the electronic scatterings and band topology that should be altered by the chirality character of the corresponding wavefunction.[13,29]

Here, by high-resolution scanning tunneling microscopy/spectroscopy (STM/STS) measurements, we demonstrate the chiral nature of the CDW phase in monolayer TiTe$_2$, characteristic of distinct triple-$q$ intensities and mirrored chiral domains. This is in excellent consistency with our density functional theory (DFT) calculations based on an extended triple-$q$ theory. Furthermore, we visualize a strong suppression of electron backscattering of Te-5$p$ hole-bands in the chiral CDW state of monolayer TiTe$_2$ from energy-resolved and layer-dependent quasiparticle interference (QPI) patterns. With the assistance of theoretical calculations, we demonstrate that the backscattering suppression stems from a chirality-driven unconventional orbital texture of electron wavefunctions, analogous to the orbital-momentum locking in DNA-type chiral molecules in the chiral CDW phase.[30] Our findings open a window for understanding the chiral nature of electron charge at the 2D limit and release the possibility for developing chiral CDW based electronic devices from an orbital degree of freedom.

Our experiments have been carried out on crystalline 1$T$-TiTe$_2$ thin films prepared by molecular beam epitaxy (MBE) on bilayer-graphene-terminated 6$H$-SiC (0001) substrates, as sketched in Figure 1a (see Methods in the Supporting Information for details). In high-temperature normal state, the TiTe$_2$ crystal structure with the $D_{3d}$ point group symmetry consists of Te-Ti-Te sandwiched sheets bonded via weak van der Waals forces, whereas within one Te-Ti-Te triple layer (TL) the hexagonal close-packed Ti atoms are octahedrally coordinated by six Te atoms. In our work, high-quality monolayer (1 TL) and bilayer (2 TL) TiTe$_2$ films with few non-stoichiometric defects have been successfully grown and identified, although self-intercalated Ti$_{1+x}$Te$_2$ multilayers invariably develop even under the extremely Te-rich growth conditions (Figures S1, S2).

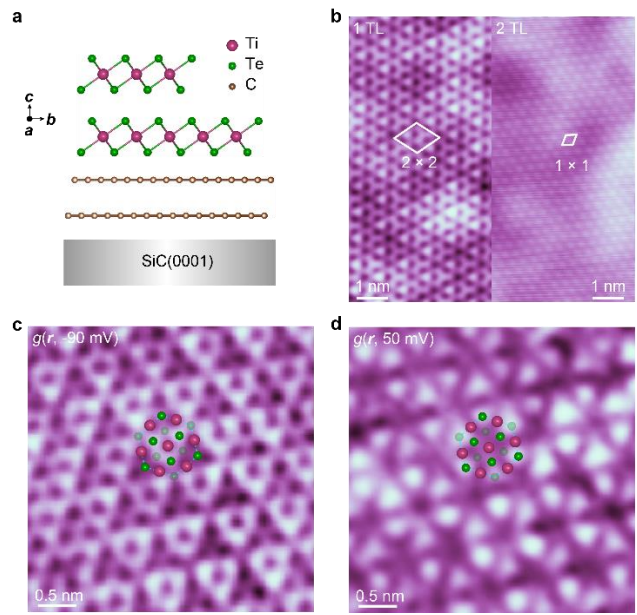

**Figure 1.** CDW in 1 TL TiTe$_2$. (a) Schematic side view from [100] direction of epitaxial 1$T$-TiTe$_2$ thin films on bilayer-graphene-terminated 6$H$-SiC(0001) substrate. (b) Typical STM topographies (5 nm × 10 nm) of 1 TL (left) and 2 TL (right) TiTe$_2$. Setpoint: $V = 50$ mV, $I = 1$ nA (1 TL); $V = 100$ mV, $I = 400$ pA (2 TL). The white rhombi mark the 2 × 2 and 1 × 1 unit cell, respectively. (c, d) Filled- and empty-state d$I$/d$V$ maps (4 nm × 4 nm) measured simultaneously in 1 TL TiTe$_2$. The opaque and semi-transparent green spheres mark the Te$_{up}$ and Te$_{down}$ atoms, respectively, while the purple spheres and dashed hexagons denote the Te atoms and CDW unit cells.

Figure 1b typifies the atomically resolved STM topographies of 1 TL and 2 TL TiTe$_2$ films, respectively, a direct comparison of which reveals the emergence of a 2 × 2 CDW order in 1 TL TiTe$_2$ and its absence in 2 TL TiTe$_2$, consistent with recent reports.[21,22] The CDW ground state of 1 TL TiTe$_2$ is further corroborated by imaging the contrast reversal of the d$I$/d$V$ maps at opposite bias polarities (Figure. 1c,d), which resemble closely those in other CDW materials 1$T$-TiSe$_2$ and 1$T$-ZrSe$_2$.[31,32] For bilayer or multilayer films, though CDW is absent in stoichiometric TiTe$_2$, a 2 × 2 superstructure exists ubiquitously in Ti$_{1+x}$Te$_2$ due to the ordering of intercalated Ti (Figure S2), which has ever been controversially ascribed to the CDW order.[33,34]

In the fast Fourier-transform (FFT) image of STM topography, the CDW modulation contributes to additional scattering spots ($q_{CDW}$) along the crystallographic directions ($q_{Bragg}$) at $q_{CDW}$ = 1/λ $q_{Bragg}$, here λ is the CDW periodicity. As reported previously, $q_{CDW}$ intensity correlates to the amplitude of the CDW modulation and would increase in a rotational way if chirality further develops.[6-8,10-13] The CDW in 1 TL TiTe$_2$ fits this case perfectly, as demonstrated in Figure 2a. For the orange- and green-enclosed regions (Figure 2b,c), the 2 × 2 CDW spots increase in intensity clockwise ($q_1$→$q_2$→$q_3$) and anticlockwise ($q_1$→$q_3$→$q_2$), respectively. The STM topography taken at the opposite bias polarity within this area reveals the same chirality (Figure S3). The robustness of the CDW chirality in 1 TL TiTe$_2$ against the sample bias is further confirmed in another region (Figure. S4).

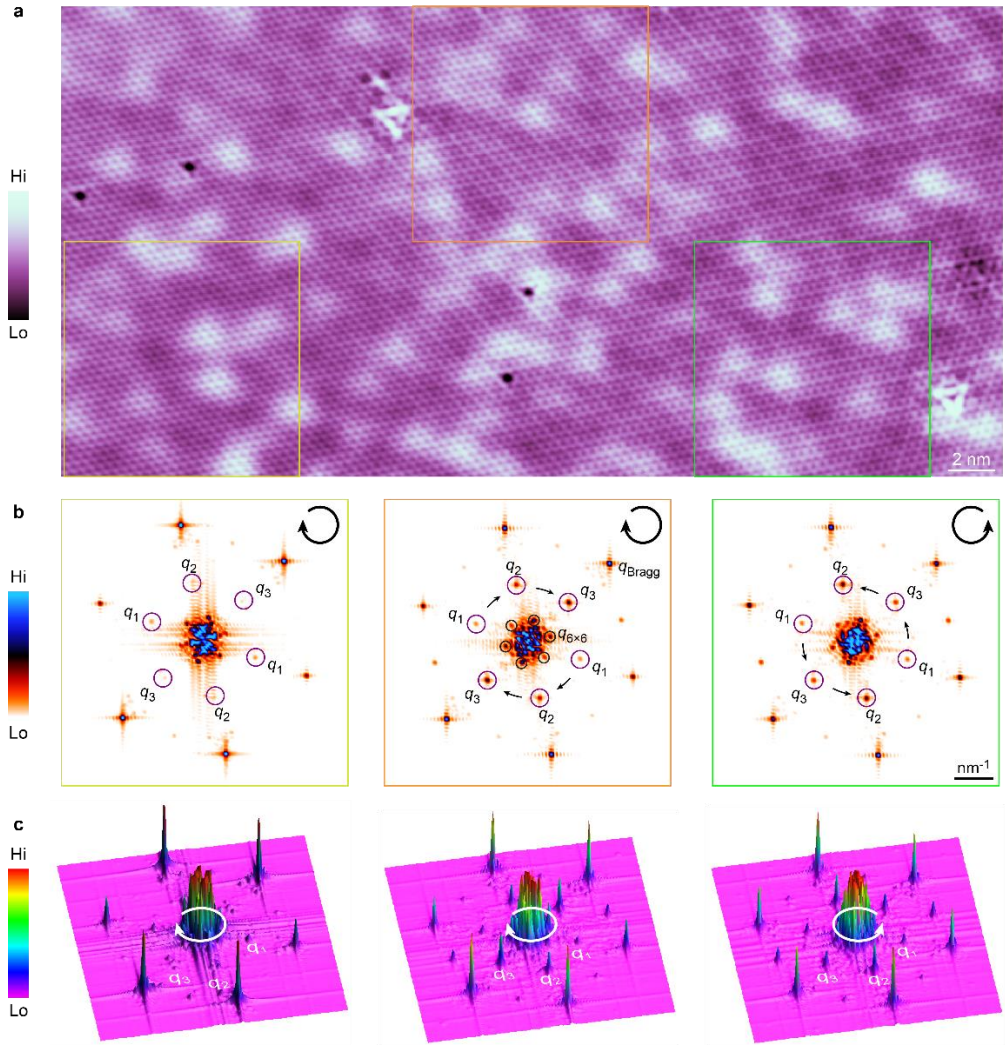

**Figure 2.** CDW chirality. (a) STM topography (40 nm × 20 nm, $V$ = 50 mV, $I$ = 1 nA) involving three chiral CDW domains as illustrated by green, orange and yellow squares. (b, c) 2D and 3D plot of the FFT images transformed from the enclosed areas by the yellow (left), orange (middle) and green (right) squares in a, with CDW intensity increasing clockwise or anticlockwise (arrowed circles). $q_{1,2,3}$ and $q_{Bragg}$ denote the CDW and Bragg peaks, while the six inner peaks ($q_{6\times6}$) stem from the 6 × 6 superstructure of underlying substrates.

The chirality induces an anisotropy of the triple-$q$ CDW vectors, lowering the threefold symmetry of the triangular lattice into a twofold one, accompanied by a mirror symmetry breaking. However, in contrast to kagome superconductor $KV_3Sb_5$,[11-13] time-reversal symmetry is preserved in 1 TL $TiTe_2$ because the CDW chirality is essentially unchanged with the orientation of the *c*-axis magnetic field (Figure S5). We also note the chiral CDW persists robustly in 1 TL $TiTe_2$ even when the charge modulation becomes substantially weak as a spatial inhomogeneity or thermal fluctuations at 78 K, as shown in Figure 2b (yellow-enclosed region) and Figure S6. Altogether, our experiments demonstrate a robust chiral CDW state in the monolayer limit of 1*T*-$TiTe_2$. This is in sharp contrast to its sister compound 1*T*-$TiSe_2$, which is the first discovered crystal hosting a chiral-CDW, yet not in the 2D limit.[5, 35, 36]

To give a microscopic understanding on the nature of chiral CDW in 1 TL $TiTe_2$, we performed first-principles calculations on the electronic structure and phonon dispersion of $TiTe_2$ (see Methods). Our results firstly show a signature of soften phonon mode in 1 TL $TiTe_2$ other than 2 TL or multilayers (Figure 3a, left and middle panels) considering a pristine structure of $TiTe_2$ without any strain ($\varepsilon = 0$). When an extension strain is switched on (Figure 3a, right panel), the phonon mode at the M point further softens and consequently favors a commensurate $2 \times 2$ CDW state in 1 TL $TiTe_2$. In Figure 3b, one can see that a minor strain of $\varepsilon \sim 3\%$ is enough to drive the CDW transition in 1 TL $TiTe_2$, but a larger lattice extension is required to induce the same CDW transition in 2 TL or bulk $TiTe_2$. In our films, strain could stem from the $TiTe_2$/graphene heterojunction interface because of the lattice mismatch and should be strongest in the monolayer limit, naturally explaining the emergent CDW in 1 TL $TiTe_2$.

Theoretically, it should be noted that the three imaginary phonon modes at M will give rise to three anisotropic $q$-vectors ($q_1$(0, 0.5), $q_2$(-0.5, 0), $q_3$(0.5, -0.5)) and corresponding displacement vectors $\vec{d}_{q,i}$, here $i$ = 1, 2, 3 denotes the atom index in a 2 × 2 supercell and $q \in \{q_1, q_2, q_3\}$. A simple superposition of these displacement vectors $\vec{d}_i = \Sigma_{q=q_{1,2,3}} \vec{d}_{q,i}$ gives a triple-$q$ combined distorted structure,[37] as sketched in Figure 3c (left panel). Such a 2 × 2 supercell structure generates an achiral CDW pattern because of the preserved $C_3$ symmetry together with the inversion symmetry.

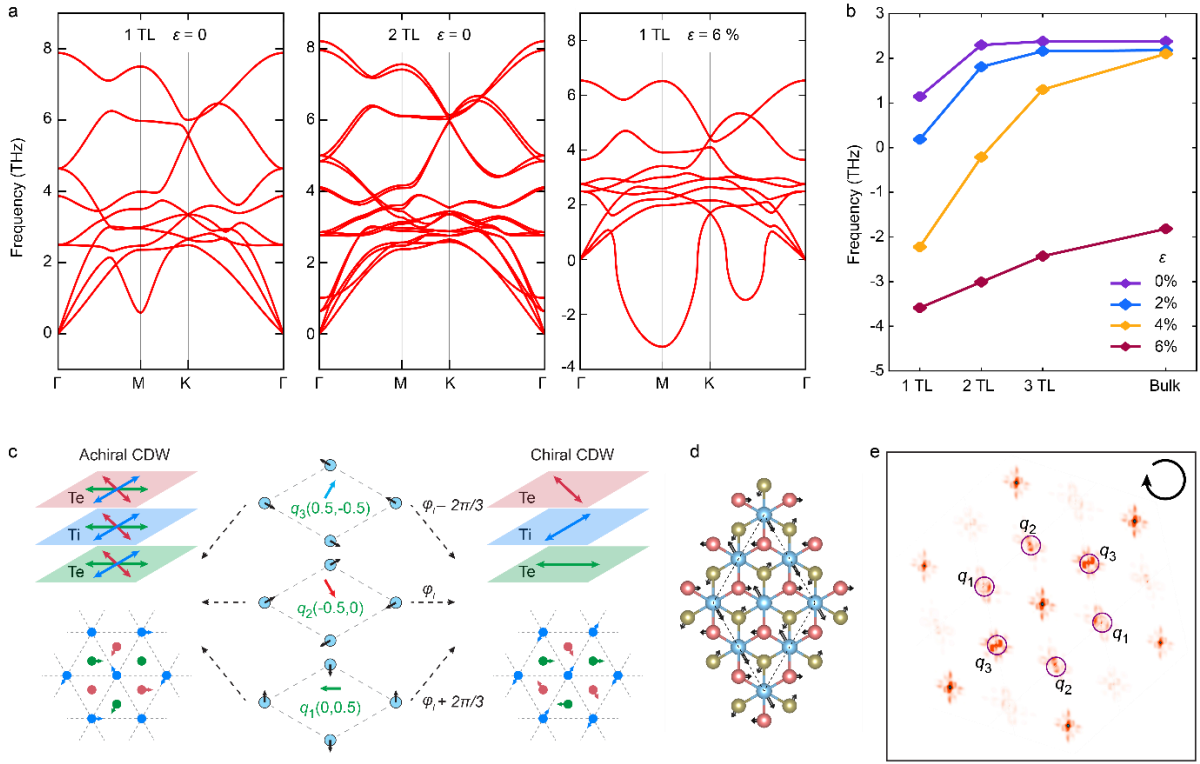

**Figure 3.** Calculated CDW instibility and its chirality. (a) Phonon dispersion of 1 TL and 2 TL TiTe$_2$ under different tensile strain $\varepsilon$ as labelled. (b) The lowest phonon frequency at M as a function of the film thickness and $\varepsilon$. (c) Schematic representation of achiral and chiral CDW. (d) Schematic of the crystal structure of 1 TL TiTe$_2$ in the chiral CDW phase. The black arrows denote the atom displacements. (e) Simulated FFT image of a chiral CDW domain (clockwise) considering a triple-$q$ CDW and a relevative phase of 2π/3.

To reproduce the chiral CDW theoretically, we broke the $C_3$ symmetry by assigning the three imaginary phonon modes at M to each $Te_{up}$-Ti-$Te_{down}$ sublayer,[19] as shown in the right panel of Figure 3c. This is achieved by adding an additional relative phase $\varphi_l \in \{0, \pm 2\pi/3\}$ to $\vec{d}_i$, but keeping the same amplitude within each sublayer, namely $\vec{d}_i = \sum_{q=q_{1,2,3}} \vec{d}_{q,i} \sin(\varphi_l + \varphi_q + \varphi_0)$, here $\varphi_l$ is the layer-dependent phase, $\varphi_q$ is the relative phase between three $q$ vectors and $\varphi_0 = \pi/2$ is the initial phase. Following the above improvements, we computed the supercell structure of 1 TL $TiTe_2$ in the chiral CDW phase with a strain of $\varepsilon \sim 6\%$ (to stabilize the CDW state) and the result is shown in Figure 3d. By employing a slightly distorted structure as an initial input, the atomic displacements of Ti and Te atoms are relaxed to be approximately $\Delta_{Ti} \sim 0.06$ Å and $\Delta_{Te} \sim 0.006$ Å in 1 TL $TiTe_2$. Besides, our calculations reveal that the top and bottom Te atoms hold the opposite phases, which means their movements in the opposite direction. The subtle atomic displacements are hardly resolved in our STM measurements within the experimental uncertainty and needs further clarification by other techniques.

From the structure in Figure 3d, we calculate the partial charge density at varied energies (Fig. S7). The broken $C_3$ symmetry enables three distinctive CDW wave vectors ($q_1$, $q_2$, $q_3$) and their stacking sequence in the $Te_{up}$-Ti-$Te_{down}$ triple-layer generates, for instance, a clockwise chiral CDW pattern in Figure 3e, in consistency with our STM measurements. Our calculations not only associate the cyclic intensity in STM with a chiral CDW but also favor that the chiral CDW in 1 TL $TiTe_2$ could originate purely from the superposition of soft phonon modes and thus can persists at the monolayer limit.

The distinct behavior of 1 TL (chiral CDW) and 2 TL (non-CDW) $TiTe_2$ enables us a comparative study on the impacts of chiral CDW on the electronic scatterings and non-trivial band topology. Figure 4a,b shows the d$I$/d$V$($r$, $E$) maps of 1 TL and 2 TL TiTe$_2$ in a 40 nm-square field of view with varied energies. One can see clearly the QPI patterns with a periodicity of 2 ~ 4 nm in the entire surface of 2 TL TiTe$_2$ regardless of the energy. However, in the 1 TL film, these QPI patterns are strongly suppressed at high-energies ($E \geq$ -0.2 eV) and faintly emerge only around native defects with decreasing energy ($E <$ -0.2 eV). Fourier transform of the d$I$/d$V$ maps into the momentum space allows a direct access to the scattering wave vectors $q$. Figure 4c,d displays the obtained FFT images at representative energies of -0.1 eV for 1 TL and 2 TL TiTe$_2$, respectively (see Figure S8 for additional FFT images under other energies). These QPI patterns in 2 TL TiTe$_2$ contribute to three ring-like scattering features (one inner ring $q_{h1}$ and two outer hexagonally warped rings $q_{h2}$ and $q_{h3}$) at the center of Brillouin zone, which vanish completely in 1 TL TiTe$_2$.

Band topology is a fundamental premise to understand the QPI patterns. As reported previously, the energy bands of TiTe$_2$ near $E_F$ are composed of a Ti 3$d$-derived electron band ($\gamma$) and two Te 5$p$-derived hole bands (inner α and outer β).[21-28] The γ band lies just below $E_F$, thus only the α and β hole bands participate in the constant energy contour (CEC) when $E \leq$ -0.1 eV, contributing two-hole pockets at Γ as sketched in Figure 4e inset. The scattering features $q_{h1}$, $q_{h2}$, and $q_{h3}$ can be easily assigned to the intra-pocket scatterings of α ($q_{h1}$) and β ($q_{h3}$) bands as well as the inter-pocket scattering between them ($q_{h2}$). Such assignment is further confirmed by the calculated dispersions of α and β bands (Figure 4e) based on the energy-dependent FFT images (some typical images are shown in fig. S8), in which the scattering geometry $q_{h1} = 2k_\alpha$ and $q_{h3} = 2k_\beta$ are considered. In our analysis, we determine the $q$ vector directly from the peak position in the FFT intensity linecut but neglect its error bar associated with the peak width for simplicity. They consist well with that from the first-principles calculations and ARPES measurements.[21-28]

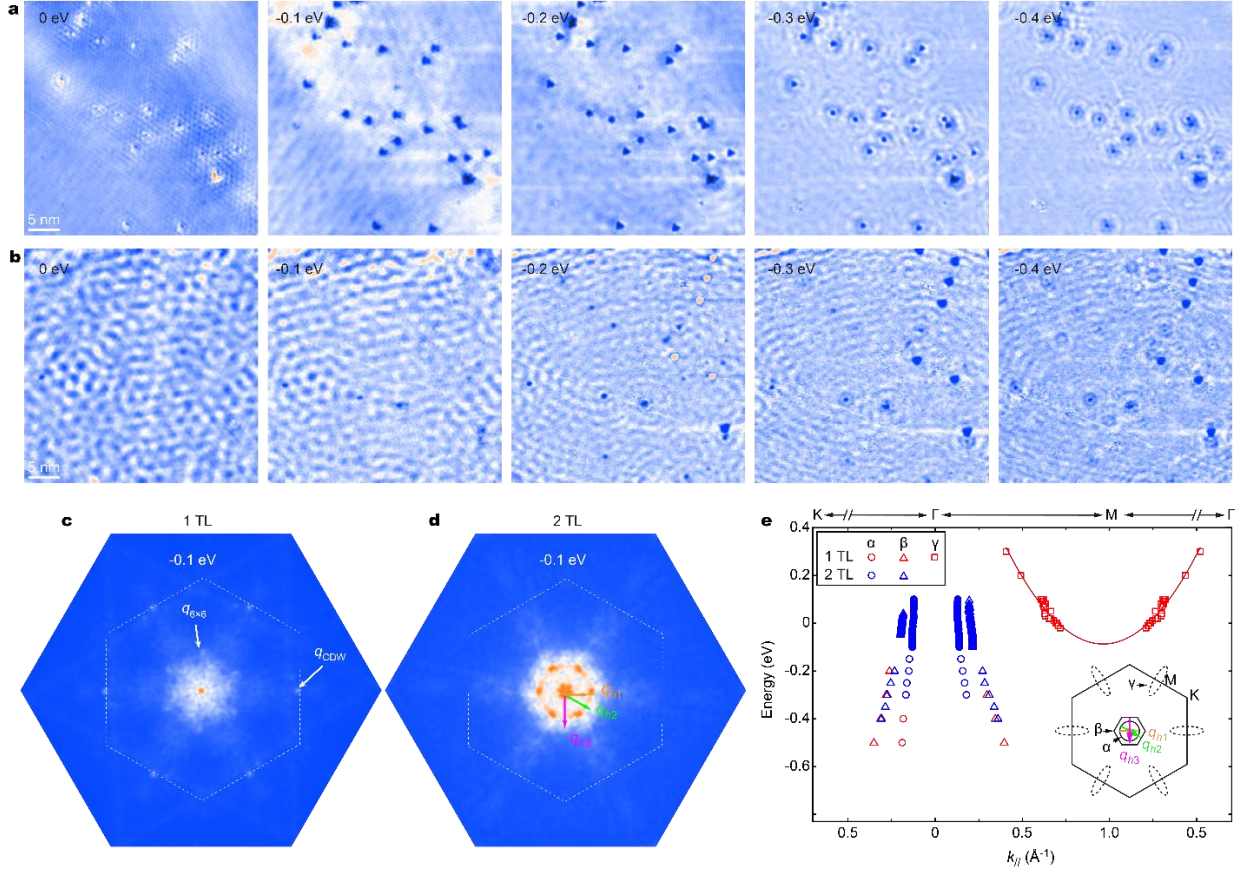

**Figure 4.** Supressed hole-band scatterings in 1 TL TiTe$_2$. (a, b) Energy-dependent d$I$/d$V$ maps (40 nm × 40 nm) measured on 1 TL and 2 TL TiTe$_2$, respectively. (c, d) FFT images at $E$ = - 0.1 eV of 1 TL and 2 TL TiTe$_2$. The white hexagons sketch out the unfolded BZ of TiTe$_2$ in the $k_z$ = 0 plane. The three wave vectors $q_{h1}$, $q_{h2}$ and $q_{h3}$ mark the scatterings from α and β bands. (e) Energy dispersions of α (circles), β (triangles) and γ (squares) bands. The bands for the 1 TL and 2 TL TiTe$_2$ are colored in red and blue, respectively. The γ band of 1 TL TiTe$_2$ is symmetrized with respect to M point and the red line is a parabolic fit to it. Inset shows the CEC of TiTe$_2$ at $E$ = - 0.1 eV, showing α and β bands and their scatterings. The dashed ellipses sketch the γ electron pocket which vanishes when $E$ = - 0.1 eV but appears in CEC at $E$ = $E_F$.

Having clarified the origin of $q_{h1}$, $q_{h2}$, and $q_{h3}$, we realize that the missing of standing waves in 1 TL TiTe$_2$ signifies a strong suppression or forbidden of the hole-band electron backscattering. This suppression, together with the STM tunneling matrix effect, result in a layer-selective band scattering of the Ti-3$d$ derived electron band ($\gamma$) and Te-5$p$ derived hole bands ($\alpha$, $\beta$) at $E_F$ when the $\gamma$ band comes into play (Figure. S8c). Considering that previous ARPES measurements cannot observe any obvious spin splitting of the Te-5$p$ bands albeit violated inversion symmetry in the chiral CDW state,[21] the absence or suppression of $q_{h1}$, $q_{h2}$, $q_{h3}$ in 1 TL TiTe$_2$ cannot be simply attributed to a helical spin texture. Besides, we further state that the backscattering suppression is not an interfacial effect from underlying substrate. Though the interfacial interaction can slightly renormalize the energy band[22], the basic band structure of the electron- and hole-bands remains robust and thus their backscatterings are expected to maintain.

Instead, by first principle calculations, we demonstrate the chirality-driven unusual orbital polarization/texture of $\alpha$ and $\beta$ bands as the microscopic origin for the suppression of their scatterings. In sharp contrast to the achiral CDW (Figure 5a,b) with negligible orbital polarization $L_z$, because of the broken inversion symmetry and preserved time-reversal symmetry ($\mathcal{T}$) in the chiral CDW state, a pronounced $L_z$ of Te-5$p$ bands can exist except for the $\mathcal{T}$-invariant points, exhibiting opposite signs at the $k$ and $-k$ wavevectors. As shown in Figure 5c,d, the backscattering between states with opposite $k$ vectors carrying opposite angular momenta is thus prohibited since $\langle -\mathbf{k}, +1|U| \mathbf{k}, -1\rangle = - \langle \mathbf{k}, -1| \mathcal{T}^{+} U \mathcal{T} | -\mathbf{k}, +1\rangle^{*} = - \langle -\mathbf{k}, +1|U| \mathbf{k}, -1\rangle = 0$, where $\mathcal{T}$ and $U$ respectively denote the time-reversal operator and time-reversal invariant operator, and $\pm 1$ denotes the angular momentum. Because different chiral CDW domains with opposite chirality in 1 TL TiTe$_2$ display an inverted orbital polarization in $k$-space, they contribute an equivalent QPI pattern with a universal suppression of hole-band backscattering.

Figure 5c,d also illustrates a much larger orbital polarization of the inner hole-band than that of the outer one. Considering the magnitude of $L_z$ difference, this would result in a stronger suppression of the scattering $q_{h1}$, followed by $q_{h3}$ and then $q_{h2}$. This prediction qualitatively consists with our experimental observations shown in Figure S6a (e.g. $E = -0.2$ eV), where $q_{h3}$ begins to show but $q_{h1}$ is still invisible. Besides, the orbital polarization of both hole-bands weakens with decreasing energy, consistent with the emergent QPI features around impurities in the FFT images. Overall, our theoretical results are in good agreement with the experimental observations of the suppression of hole-pocket scatterings, indicating the chirality-driven unconventional orbital texture in the chiral CDW phase.

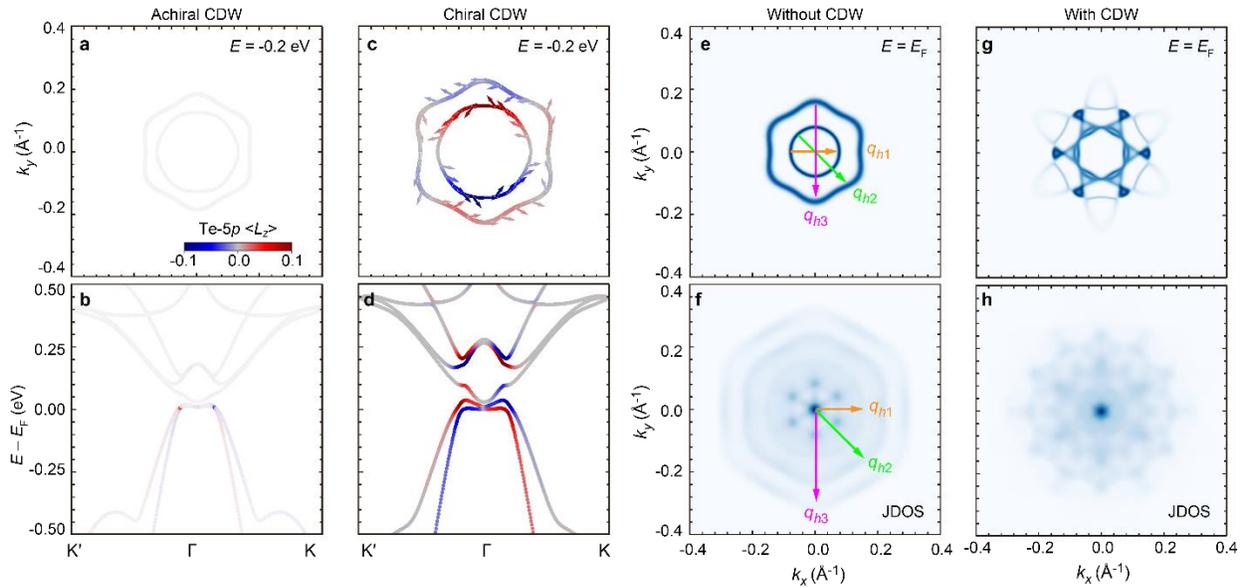

**Figure 5.** Backscattering-immune orbital texture. (a, c) Calculated CEC at - 0.2 eV of 1 TL TiTe$_2$ in a achiral and a chiral CDW state. In our calculations, we use the structure shown in Fig. 3c-d as the input and spin-orbit coupling is not included. Arrows and colors of the vectors represent the in-plane ($L_x$, $L_y$) and out-of-plane ($L_z$) components of Te-5$p$ orbital angular momentum, respectively. (b, d) Band structure along the K'-Γ-K direction with projected orbital

polarization $L_z$ in a achiral and a chiral CDW state. (e, g) Calculated CEC at $E_F$ around the BZ center from the effective model without and with a CDW distortion. Chirality is not included. (f, h) Corresponding JDOS calculated by $JDOS(q, E) = \int I(k, E)I(k + q, E)d^2k$, here $I(k, E)$ is the delta function from the charge densities.

It is worth mentioning that a momentum-dependent hybridization effect in consequence of the distinct orbital characters of the inner and outer hole bands,[27, 28] also helps suppress the hole-band backscattering. Here, we adopt a low-energy effective model (see Methods) to capture the Te-5$p$ states derived snowflake QPI pattern at the BZ center of $E_F$ (Fig. S8). Figure 5e displays the computed Fermi surface contour of hole bands without the CDW transition. In this normal state ($T > T_{CDW}$), the two hole-bands from Te-5$p$ orbitals contribute three hole-pocket scatterings as anticipated (Figure 5f). As the CDW distortion is turned on ($T < T_{CDW}$), the backfolded Ti-3$d$ electron-bands strongly hybridize to the outer valence bands, modifying the band structure (Figure. S9) and Fermi surface contour dramatically (Figure 5g). A bunch of new formed small pockets combined with enhanced Ti-$d$ and Te-$p$ orbital charge transfer in the CDW state give rise to a significant quench of the original hole pocket scattering, as revealed by the joint density of states (JDOS) (Figure 5h). Such a strong band-warping effect is not captured by DFT. It should be noted that this effect could only occur within the hybridization gap, which is estimated to be smaller than 100 meV in the low-energy d$I$/d$V$ spectra (Figure S10). Such a small band hybridization gap is insufficient to explain the observed backscattering suppression at higher energy, such as $E = -0.2$ eV. After invoking chirality, both the backscattering suppression and its energy-range (from $E_F$ to -0.4 eV) can be well explained by the chirality induced orbital polarization

Our findings report the first direct visualization of chiral CDW in the monolayer limit of TiTe$_2$, in which the dimensionality, i.e., the ultrathin thickness, still plays a crucial role in the formation of chirality. The spontaneously formed chiral CDW state in monolayer TiTe$_2$ arises from a relative phase difference between the triple-$q$ CDW vectors embedded in each atomic layer, which is in sharp contrast to the unparalleled $q_{Bragg}$ and $q_{CDW}$ induced chiral Fermi surface, such as in TaS$_2$.[29] Such a helix charge order in monolayer TiTe$_2$ is thus expected to enable plentiful physical responses such as chirality-induced spin selectivity[30] and exotic circular dichroism,[38] as exemplified by a recent work on indium nanowires.[39]

Furthermore, our work unveils an underlying relationship between the suppression of Fermi-pocket scatterings and the chirality-driven orbital texture, indicating that the orbital angular momentum plays a significant role in these phenomena. Unlike Rashba spin textures in topological insulators,[40,41] we provide a novel insight that the backscattering of low-energy electrons that determines the transport behavior can be unusually dominated by an orbital degree of freedom. Spectroscopic visualization of the electronic scattering processes offers a direct way to explore such orbital polarization effect. Considering the CDW distortion and substrate effect, it is undeniable that the inclusion of spin-orbit coupling may quench the polarization at certain $k$ points, which requires further circular dichroism measurements or other investigations. In this regard, the chiral CDW order can be desirably tuned and optimized by a range of manners such as doping,[23] pressure and Moiré pattern[25] to broaden the potential applications of the chiraltronics.

## ASSOCIATED CONTENT

**Supporting Information:** The Supporting Information is available free of charge online.

Methods; Identification of stoichiometric TiTe$_2$ films; CDW chirality; Suppressed hole-pocket backscattering in 1 TL TiTe$_2$.

AUTHOR INFORMATION

**Author contributions:** C.L.S., X.C.M. and Q.K.X. conceived and designed the experiments. M. Q. R., F.J.C., Q.J.C carried out the MBE growth and STM measurements. M.Q.R. and C.L.S. analyzed the experimental data. Y.F.Z., B.H.Y. and Q.H.L. carried out the theoretical calculations. M.Q.R., Y.F.Z., Q.H.L. and C.L.S. wrote the manuscript with comments from all authors. M.Q.R., F.J.C., and Y.F.Z. contributed equally to this work.

**Notes:**

The authors declare no competing financial interest.

ACKNOWLEDGMENT

The work was financially supported by the National Key R&D Program of China (Grants No. 2022YFA1403100), the Natural Science Foundation of China (Grants No. 52388201, No. 62074092). Q.H.L. acknowledge the support of Center for Computational Science and Engineering of Southern University of Science and Technology.

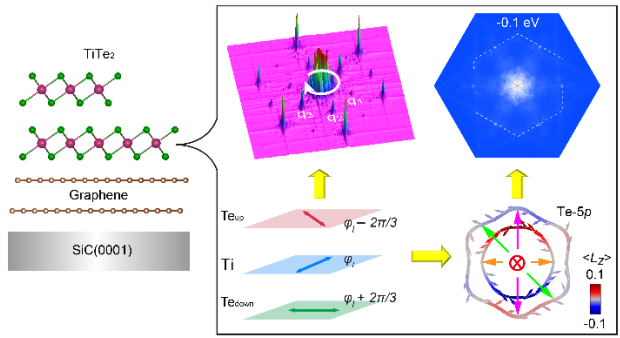

TOC